\documentclass[a4paper,twocolumn,aps,prl]{revtex4}

\newcommand{\eq}[1]{\begin{equation}#1\end{equation}}
\newcommand{\dd}{\mathrm{d}}
\newcommand{\ee}{\mathrm{e}}

\newcommand{\Ai}{\mathrm{Ai}}
\newcommand{\Tr}{\mathrm{Tr \,}}
\newcommand{\rp}{\mathrm{Re\,}}
\newcommand{\ip}{\mathrm{Im \,}}

\newcommand{\identity}{\openone}

\newcommand{\mupr}{|\Uparrow \, \rangle}
\newcommand{\mupl}{\langle \, \Uparrow|}

\newcommand{\mnt}{\mathcal{M}_n(t)}
\newcommand{\jwr}{|\mathrm{JW}\rangle}
\newcommand{\jwl}{\langle \mathrm{JW} |}

\newcommand{\rnr}{|0\rangle_{\mathrm{R}}}
\newcommand{\nsnr}{|0\rangle_{\mathrm{NS}}}

\newcommand{\nsr}[1]{|#1\rangle_{\mathrm{NS}}}
\newcommand{\nsl}[1]{{_\mathrm{NS}}\langle #1 |}
\newcommand{\rr}[1]{|#1\rangle_{\mathrm{R}}}
\newcommand{\rl}[1]{{_\mathrm{R}}\langle #1 |}

\usepackage{amsmath}
\usepackage{amsfonts}
\usepackage{graphics}
\usepackage{graphicx}
\usepackage{psfrag}
\usepackage{color}
\usepackage{colordvi}
\usepackage{bbm}

\begin{document}

\title{Hydrodynamical phase transition for domain-wall melting in the XY chain}

\author{Viktor Eisler and Florian Maislinger}
\affiliation{
Institut f\"ur Theoretische Physik, Technische Universit\"at Graz, Petersgasse 16,
A-8010 Graz, Austria
}

\begin{abstract}
We study the melting of a domain wall, prepared as a certain low-energy excitation above
the ferromagnetic ground state of the XY chain. In a well defined parameter regime
the time-evolved magnetization profile develops sharp kink-like structures in the bulk,
showing features of a phase transition in the hydrodynamic scaling limit.
The transition is of purely dynamical nature and can be attributed to the appearance
of a negative effective mass term in the dispersion. The signatures are also clearly visible
in the entanglement profile measured along the front region, which can be obtained by
covariance-matrix methods despite the state being non-Gaussian.
\end{abstract}

\maketitle

Uncovering the mechanism of phase transitions belongs to one of the
most spectacular achievements of statistical physics.
The abrupt changes in the properties of matter, in response
to the tuning of a control parameter, could be understood through simple concepts such as
order parameter, symmetry breaking, or free energy. 
While the theory is well established for systems in thermal equilibrium,
and can even be extended to quantum phase transitions at zero temperature \cite{SS},
it is far from obvious how these concepts generalize to the nonequilibrium scenario.

Due to this ambiguity, there has been various attempts
to lift the definition of a phase transition into the dynamical regime.
In the particular context of quantum quenches \cite{PSSV11,CEM16},
dynamical quantum phase transitions (DQPT)
were introduced by analogy, via the definition of a dynamical free energy density \cite{HPK13}.
It is simply given via the overlap between initial and time-evolved states, and 
DQPT manifests itself in the nonanalytic real-time behavior of this return probability,
see \cite{Heyl17} for a recent review. Despite not being a conventional observable,
the return probability and the signatures of a DQPT could directly be detected
in a recent experiment \cite{Jurcevic17}.

On the other hand, in a number of approaches the definition of dynamical
phases is based on the time-asymptotic behavior of an order parameter that
shows abrupt changes when crossing the phase boundaries.
Dynamical phase transitions based on a suitable order parameter have been
identified for quench protocols of various closed many-body systems \cite{EKW09,SF10,SB10}
and  the studies have even been extended to the open-system scenario \cite{GL10,DTMFZ10}.
Furthermore, connections between the different concepts of a DQPT, based on dynamical
free energy vs. order parameter, have recently been pointed out \cite{HZ17,ZHKS18}.

Here we shall address the question whether a phase transition in simple quantum chains
might occur due to the presence of initial spin gradients, which drive the system
towards a nonequilibrium steady state (NESS).
In the context of Markovian open system dynamics, such an example was found
earlier for a boundary driven open XY spin chain, where the emergence of long range order
was observed in the NESS below a critical value $h<h_c$ of a model parameter \cite{PP08}.
Although the phenomenon seems robust enough against the details of incoherent driving \cite{PZ10},
no counterpart of the phase transition under closed unitary dynamics has been found so far.

\begin{figure}[t]
\center
\includegraphics[width=\columnwidth]{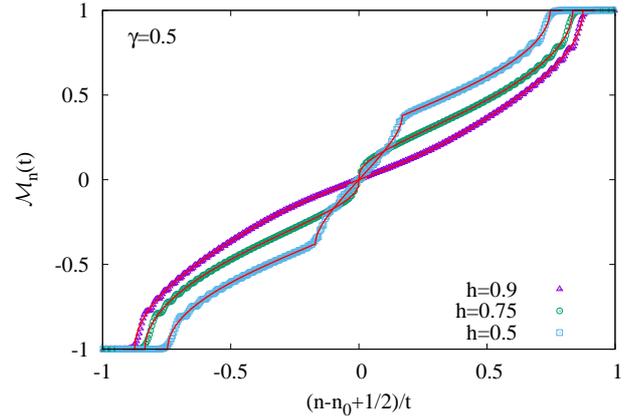}
\caption{Normalized magnetization profiles (symbols) at $t=200$ compared to the
hydrodynamic solution (red solid lines) in \eqref{mntsc}. 
The phase transition is located at $h_c=1-\gamma^2=0.75$.}
\label{fig:mag}
\end{figure}

To mimic the effect of gradients imposed at the boundaries in the open system setup,
here we prepare instead a domain-wall initial state and then let the system evolve under its
own unitary dynamics. The domain wall is created as a simple low-lying excitation
above the ferromegnatic (symmetry-broken) ground state of the XY chain.
Our main result is illustrated on Fig. \ref{fig:mag}, where the qualitative change in the
time-evolved and properly normalized magnetization profiles is clearly visible.
The phase transition point $h_c$ exactly coincides with the one found in Ref. \cite{PP08},
and is signalled by an infinite slope in the center of the profile, whereas kinks are
developing in the bulk for $h<h_c$. The nonanalytical behavior appears only in the
hydrodynamical limit, shown by the solid lines in Fig. \ref{fig:mag}.
However, in contrast to Ref. \cite{PP08}, our results on the correlations indicate that
the NESS itself is similar to the symmetry-restored ground state of the chain and does not
show any criticality around $h_c$. Hence we use the term hydrodynamical phase
transition to distinguish between the two behaviors.


%
%

The Hamiltonian of the XY chain is given by
\eq{
H=- \sum_{n=1}^{N-1} \left( \frac{1+\gamma}{4}\sigma_n^x 
\sigma_{n+1}^x+\frac{1-\gamma}{4}\sigma_n^y \sigma_{n+1}^y\right)
-\frac{h}{2} \sum_{n=1}^{N} \sigma_n^z
\label{hxy}}
where $\sigma^\alpha_n$ 
are Pauli matrices on site $n$, $\gamma$ is the anisotropy and
$h$ is a transverse magnetic field. The XY model can be mapped to a chain
of free fermions via a Jordan-Wigner (JW) transformation, by introducing the
Majorana operators 
\eq{
a_{2j-1} = \prod_{k=1}^{j-1} \sigma^z_k \sigma^x_{j}, \qquad
a_{2j} = \prod_{k=1}^{j-1} \sigma^z_k \sigma^y_{j},
\label{maj}}
satisfying anticommutation relations $\left\{a_k,a_l\right\}=2\delta_{k,l}$.
While the open boundaries in Eq. \eqref{hxy} are most suitable for
numerical investigations of the dynamics on finite size chains, for the analytical treatment
one should impose antiperiodic boundary conditions
$\sigma^{x,y}_{N+1}=-\sigma^{x,y}_1$ on the spins,
such that $H$ can be brought into a diagonal form by
a Fourier transform and a Bogoliubov rotation \cite{sm}.

We focus on the
parameter regime $0<\gamma\le 1$ and $0 \le h<1$, where the model
is in a gapped ferromagnetic phase, with magnetic order in the $x$ direction.
In particular, in the limit $N\to \infty$,
the ground state is twofold degenerate, with $\nsr{0}$ and $\rr{0}$
located in the Neveu-Schwarz (NS) and Ramond (R) sectors, corresponding
to $\pm 1$ eigenvalues of the parity operator $P=\prod_{k=1}^N \sigma^z_k$,
which commutes with the Hamiltonian $\left[H,P\right]$=0.
Since both of the ground states are parity eigenstates, their 
magnetization is vanishing. However, starting from the symmetry-broken
ground state $\mupr$, a domain wall initial state can be prepared via a JW excitation,
i.e. acting with a single Majorana operator as
%
\eq{
\jwr = a_{2n_0-1} \mupr, \qquad
\mupr = \frac{\nsnr + \rnr}{\sqrt{2}}.
\label{jw}}
In numerical calculations we always consider domain walls localized
in the middle of the chain, $n_0=N/2+1$.

Our primary goal is to calculate the magnetization profile in the time evolved state
\eq{
\ee^{-iHt}\jwr=\frac{\nsr{\phi_t}+\rr{\phi_t}}{\sqrt{2}}
\label{tevol}}
being a superposition of states from the two parity sectors.
Both can be obtained by rewriting the excitation in \eqref{jw}
in the fermionic eigenbasis of the Hamiltonian, leading to a superposition
of single-particle states. These can then be trivially time evolved and yield \cite{sm}
\eq{
\nsr{\phi_t}= 
\frac{1}{\sqrt{N}} \sum_{q \in \mathrm{NS}}
\ee^{-i\epsilon_qt} \ee^{-iq(n_0-1)}\ee^{i\theta_q/2} \nsr{q},
\label{phit}}
where the single-particle dispersion $\epsilon_q$ and the
Bogoliubov phase $\theta_q$ are given by
\eq{
\begin{split}
&\epsilon_q = \sqrt{(\cos q-h)^2+\gamma^2 \sin^2 q} \, , 
\\
&\ee^{i(\theta_q+q)} = \frac{\cos q -h + i \gamma \sin q}{\epsilon_q} \, .
\end{split}
\label{epsqba}
}
The result for $\rr{\phi_t}$ is completely analogous to \eqref{phit},
with the sum running over momenta $p \in R$.
In turn, the normalized magnetization can be cast in the form
\eq{
\mnt=\frac{\jwl \sigma^x_n(t)\jwr}{\mupl \sigma^x_n \mupr}
= \rp \rl{\phi_t} \hat{\mathcal{M}}_n \nsr{\phi_t} \,,
\label{mnt}}
where, in the limit $N\gg1$, the form factors read \cite{Iorgov11,IL11}
\eq{
\rl{p}\hat{\mathcal{M}}_n\nsr{q} =
-\frac{i}{N}\frac{\epsilon_p + \epsilon_q}{2\sqrt{\epsilon_p \epsilon_q}}
\frac{\ee^{i(n-1/2)(q-p)}}{\sin \frac{q-p}{2}}.
\label{ff}}

Combining the results \eqref{phit}-\eqref{ff} and considering the thermodynamic limit,
one ends up with a double integral formula for the magnetization \cite{sm}.
Interestingly, this is exactly the same expression as the one found earlier for the
transverse Ising (TI) chain \cite{EME16}, except that the form of the dispersion 
and the Bogoliubov angle \eqref{epsqba} are now more general. In fact, it is the very presence
of the XY anisotropy that will give rise to a peculiar dynamical behavior.
The hydrodynamical phase transition is encoded in the $q \ll 1$
expansion of the dispersion
\eq{
\epsilon_q \approx \Delta + \frac{h-h_c}{2\Delta}q^2+c \, q^4 ,
\label{epsq0}}
where $\Delta = 1-h$ is the excitation gap and $h_c=1-\gamma^2$ is a critical field.
The coefficient $c$ has a lengthy expression in terms of $h$ and $\gamma$, satisfying
$c>0$ for any $h<h_c$. In contrast, the mass term in Eq. \eqref{epsq0} becomes
negative below the critical field.

%
\begin{figure}[th]
\center
\includegraphics[width=\columnwidth]{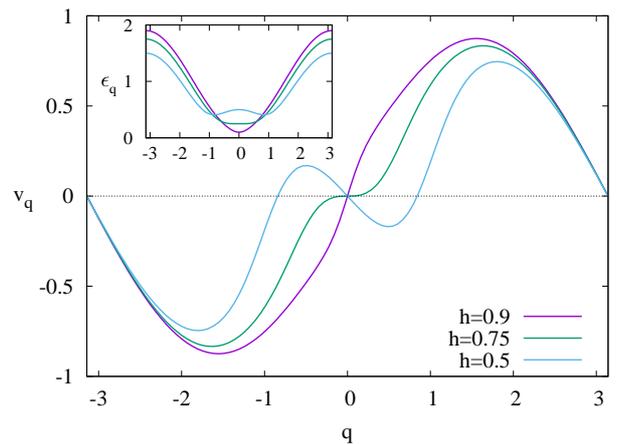}
\caption{Single-particle velocities $v_q$ and dispersion $\epsilon_q$ (inset).}
\label{fig:vq}
\end{figure}
%

While a negative effective mass has no effect on the ground-state properties,
it will play a crucial role in the dynamics. Indeed, in a well-defined limit,
the shape of the melting domain wall is entirely determined by the group velocities
$v_q=\frac{\dd \epsilon_q}{\dd q}$. These are shown on Fig. \ref{fig:vq} for $\gamma=0.5$,
and three different magnetic fields above, below and at the critical value $h_c$.
In case $h<h_c$, the negative slope of $v_q$ around $q \to 0$ leads to the development
of a new local maximum, which eventually gives rise to a nonanalytic behavior in
the hydrodynamic profiles of various observables. In particular, 
introducing the scaling variable $\nu = (n-n_0+1/2)/t$, the magnetization profile
reads
\eq{
\mnt = 
1-2\int_{-\pi}^{\pi} \frac{\dd q}{2\pi}  \Theta(v_{q} - \nu),
\label{mntsc}}
where $\Theta(x)$ is the Heaviside step function.
The result \eqref{mntsc} follows rigorously
from a stationary-phase analysis \cite{sm} of the integral representation of $\mnt$,
and has a clear physical interpretation.
Namely, each single-particle excitation carries
a spin-flip \cite{SY97,AKR08,RI11,KMZ17} and thus the magnetization along a fixed ray follows
from the integrated density of excitations whose speed exceeds $\nu$. 
Hence, for $h<h_c$ the nonanalytical behavior of the density is a consequence
of the new branch of solutions around the local maximum for negative momenta.

The comparison between the profiles and the hydrodynamic scaling function is
shown on Fig. \ref{fig:mag}. The magnetization at $t=200$ and various $h$ were
calculated for an open chain of size $N=400$
using the Pfaffian formalism described in \cite{EME16}. One has an excellent
agreement with clear signatures of the developing kink for $h<h_c$.
The hydrodynamic profile in general depends on the details of the dispersion and is
hard to obtain analytically, since the solution of $v_q=\nu$ leads to a fourth-order equation.
Nevertheless, one expects a universal behavior to emerge around the edge
of the front \cite{ER13}. Indeed, the stationary phase calculation around $v_{q_*}=v_{max}$
can be extended to capture the fine structure of the front \cite{VSDH15,ADSV16,PG17,Kormos17},
suggesting the following choice for the scaling variable
\eq{
X = (n-n_0+1/2+\theta'_{q_*}/2-v_{q_*}t) \left(\frac{2}{|v''_{q_*}|t}\right)^{1/3}.
\label{X}}
In turn, the edge magnetization is given by \cite{sm}
\eq{
\mnt = 
1-2 \left(\frac{2}{|v''_{q_*}|t}\right)^{1/3} \rho(X) \, ,
\label{medge}}
where $\rho(X)=\left[ \Ai'(X) \right]^2 - X\Ai^2(X)$ is just the diagonal part of the
Airy kernel \cite{TW94}.

%
\begin{figure}[htb]
\center
\includegraphics[width=\columnwidth]{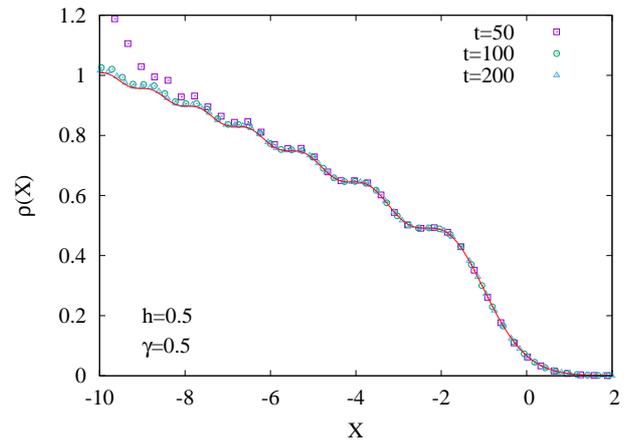}
\caption{Edge scaling of the magnetization profile, with the scaling variable $X$ and function
$\rho(X)$ defined by Eqs. \eqref{X} and \eqref{medge}, respectively.}
\label{fig:edge}
\end{figure}
%

The edge scaling \eqref{medge} is tested against numerical calculations for $h=\gamma=0.5$
in Fig. \ref{fig:edge}, showing an excellent agreement already for moderately large times.
Note that the larger deviation towards the bulk for $t=50$ is due to the presence of the kink
in the profile. In fact, one could ask whether zooming on around the kink would yield a similar
universal fine structure as for the edge. However, in the latter case the density has a
nonuniversal bulk contribution superimposed, which spoils the step structure.
It is also worth noting that the edge scaling \eqref{medge} for the XY chain can not
be derived from a simple higher-order extension of the hydrodynamical picture \cite{Fagotti17}.

The signatures of the hydrodynamical phase transition are also visible on the
entanglement profiles, as measured by the von Neumann entropy between the segment
$A=\left[1,N/2+r\right]$ and $B$ the rest of the system. Although the XY chain maps to
free fermions, extracting the entropy via covariance-matrix techniques for Gaussian states
\cite{Vidal03,PE09} requires some additional care.
Indeed, the initial state is excited from the symmetry-broken ground
state of the model, which is inherently non-Gaussian \cite{FE13}.
This difficulty can, however, be overcome by the following considerations.
Let us denote by $\rho_\Uparrow$ the reduced density matrix (RDM) arising from
the time evolved state \eqref{tevol} after tracing out the degrees of freedom in $B$.
The arrow indicates the choice of the symmetry-broken ground state in \eqref{jw}
and the entropy of the RDM is given by $S(\rho_\Uparrow)=- \Tr \rho_\Uparrow \ln \rho_\Uparrow$.
In fact, one could equally well have defined $\rho_\Downarrow$ starting from the spin-reversed initial state,
with the entropies of the two RDMs satisfying
$S(\rho_\Uparrow)=S(\rho_\Downarrow)$
due to obvious symmetry reasons. 
The main trick is now to consider the convex combination
\eq{
\rho_G = \frac{\rho_\Uparrow + \rho_\Downarrow}{2},
\label{rhog}}
which removes all the parity-odd contributions from the RDMs, albeit still mixing
parity-even terms from the two sectors $\mathrm{NS}$ and $\mathrm{R}$.
However, in the thermodynamic limit all the expectation values of local
operators become equal in both sectors \cite{FE13}, hence $\rho_G$
is equivalent to a Gaussian RDM where the excitation is created upon
the parity-symmetric ground state $\nsr{0}$. 

Due to its Gaussianity, the entropy
of $\rho_G$ can now be obtained by applying the covariance-matrix
formalism as shown in Ref. \cite{CR17}. Indeed, the effect of the Majorana
excitation can be represented in a Heisenberg picture
\eq{
a'_k = a_{2n_0-1} a_k a_{2n_0-1}=\sum_{l=1}^{2N} Q_{k,l} a_l, \qquad
}
as an orthogonal transformation on the Majoranas, with matrix
elements $Q_{k,l}=\delta_{k,l}(2\delta_{k,2n_0-1}-1)$. Similarly,
time evolving the state corresponds to the transformation
\eq{
a'_k(t) = \ee^{iHt} a'_k \ee^{-iHt}=\sum_{l=1}^{2N} R_{k,l} a'_l, \qquad
}
with matrix elements $R_{k,l}$ given as in Ref. \cite{EME16}.
Hence $\rho_G$ corresponds to a RDM associated to the Gaussian state
with covariance matrix
\eq{
\tilde \Gamma = R \, Q \, \Gamma \, Q^T R^T ,
\label{tg}}
where $i\Gamma_{k,l}=\nsl{0} a_k a_l \nsr{0}-\delta_{k,l}$. Note that the matrix
$\tilde \Gamma$ is exactly the one that appears in the Pfaffian by the
calculation of the magnetization \cite{EME16}.


Although the entropy of $\rho_G$ follows simply via the eigenvalues of the
reduced covariance matrix $\tilde \Gamma_A$ \cite{Vidal03,PE09},
one still has to relate it to the entropy of the non-Gaussian RDM $\rho_\Uparrow$
that we are interested in.
To this end, one can make use of the inequality for convex combinations of 
density matrices \cite{LR68,Wehrl78}
\eq{
S(\sum_i \lambda_i \rho_i) \le
\sum_i \lambda_i S(\rho_i) - \sum_i \lambda_i \ln \lambda_i \, .
}
Furthermore, it is also known that the inequality is saturated if the
ranges of $\rho_i$ are pairwise orthogonal. Applying it to
Eq. \eqref{rhog}, the orthogonality condition is clearly satisfied due
to $\langle \Uparrow | \Downarrow \rangle=0$ and hence one arrives at
\eq{
S(\rho_\Uparrow) = S(\rho_G) - \ln 2 \, .
\label{sup}}
The entropy can thus be exactly evaluated using Gaussian techniques.

\begin{figure}[htb]
\center
\includegraphics[width=\columnwidth]{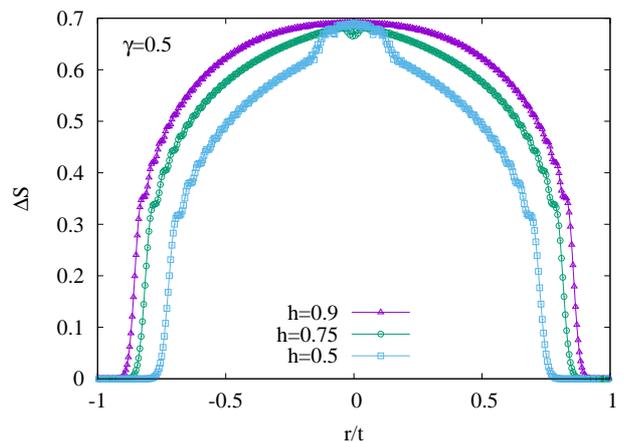}
\caption{Entanglement profiles as a function of the rescaled distance $r$
of the cut from the middle of the chain. The entropy  difference $\Delta S$
from the initial state value is shown at $t=200$ for the same parameter values as in Fig. \ref{fig:mag}.}
\label{fig:ent}
\end{figure}

The result for the profile $\Delta S$, measured from the $t=0$ value,
is shown on Fig. \ref{fig:ent} at time $t=200$, against the rescaled cut position.
The parameters are chosen to be identical to Fig. \ref{fig:mag}, and
a kink for $h=0.5$ emerges again at the value of $r/t$ equal to the local
maximum of the velocity $v_q$. Furthermore, the entropy growth for the half-chain ($r/t=0$)
clearly converges towards the value $\ln 2$, which can be interpreted as
a restoration of the spin-flip symmetry in the NESS. Note also the light dip in the
middle for $h=h_c=0.75$, which is the consequence of a much slower convergence
towards the NESS at criticality. The entropy profiles obtained by the Gaussian
technique have also been compared to the results of density-matrix renormalization group \cite{Schollwoeck11}
calculations, finding an excellent agreement and thus justifying the result in Eq. \eqref{sup}.

%
%

We finally consider the normalized equal-time spin-correlation functions
$\mathcal{C}_{m,n}(t)= \nsl{\phi_t} \hat{\mathcal{M}}_m \hat{\mathcal{M}}_n \nsr{\phi_t}$
which can be studied via the form-factor approach by inserting a resolution of the identity
between the operators. Although in general all the multi-particle form factors are nonvanishing,
the dominant contribution to the correlations comes from the single-particle terms
\eq{
\mathcal{C}_{m,n}(t)
\simeq \sum_p \nsl{\phi_t} \hat{\mathcal{M}}_m \rr{p}\,\rl{p} \hat{\mathcal{M}}_n \nsr{\phi_t} \, .
\label{cmnt}
}
The above expression can again be evaluated in the hydrodynamic scaling limit
and for $m<n$ yields \cite{sm}
%
\eq{
\mathcal{C}_{m,n}(t) \simeq
1-2\int_{-\pi}^{\pi} \frac{\dd q}{2\pi}  \Theta(v_{q} - \mu) \Theta(\nu - v_{q}) \, ,
\label{cmnt2}}
where $\mu$ is defined analogously to $\nu$. The integral in \eqref{cmnt2}
gives the number of excitations with velocities between the rays defined by
$\mu$ and $\nu$, and has again a simple interpretation.
In fact, it is directly related to the difference of the magnetizations along those
rays and thus shows similar nonanalytical behavior for $h<h_c$.

In the NESS limit $t\to\infty$ with $m,n$ fixed, Eq. \eqref{cmnt2} predicts
long-range magnetic order $\mathcal{C}_{m,n}(t) \to 1$.
Together with $\mnt \to 0$, this behavior is characteristic of the ground state
$\nsr{0}$ at large separations $n-m \gg 1$.
Furthermore, a careful numerical analysis shows that $\mathcal{C}_{m,n}(t)$
converges towards the proper ground-state value even for small separations of
the spins. Indeed, in the ferromagnetic regime the normalized correlators deviate
from unity by a term decaying exponentially with the distance \cite{FF}.
The source of the discrepancy is the approximation in \eqref{cmnt}, which
neglects the contribution of the multi-particle form factors.
A detailed analysis of the correlations will be presented elsewhere \cite{EMup}.

In conclusion, our studies of domain-wall melting in the XY chain 
have revealed a phase transition, manifest in the emergence of kinks
in the profiles of various observables.
While the critical point $h_c=1-\gamma^2$ coincides with the one
found earlier for open-system dynamics \cite{PP08}, the transition exists
only in the hydrodynamic regime, and does not survive the NESS limit.
In contrast, the latter one seems to be given by the parity-symmetric
ground state, which does not show any criticality around $h_c$.

Although demonstrated on a simple free-fermion example, there is good
reason to believe that this phenomenon carries over to generic integrable
systems, where the proper hydrodynamic description has only recently
been identified \cite{BCDNF16,CADY16} and applied to initial states
with domain walls \cite{PDNCBF17,CDLV18}.
In particular, the emergence of kinks in the magnetization profile
has been observed for the XXZ chain at large anisotropies, resulting from
the velocity maxima of the various quasiparticle families that govern the
hydrodynamics \cite{PDNCBF17}. While the mechanism seems to be closely related
to the one presented here, it is unclear whether a hydrodynamical phase
transition point exists in the XXZ case, since all the profiles considered
in \cite{PDNCBF17} belong to the kink phase.

Finally, it remains to be understood whether the finite increase of entropy
after the JW excitation could be interpreted within a framework similar to
the one introduced for local operator insertions in conformal field theories \cite{NNT14}.
While the results have been checked against the lattice equivalent
of local primary excitations for the transverse Ising chain in Ref. \cite{CR17}, it would be
interesting to see whether the field theory treatment could be generalized to include
the massive case and the non-local operators considered here.

We thank H. G. Evertz and M. Fagotti for discussions.
The authors acknowledge funding from the Austrian Science Fund (FWF) through
Project No. P30616-N36, and through SFB ViCoM F41 (Project P04).

\bibliographystyle{apsrev.bst}

\bibliography{xyfront_refs}

\begin{thebibliography}{44}
\expandafter\ifx\csname natexlab\endcsname\relax\def\natexlab#1{#1}\fi
\expandafter\ifx\csname bibnamefont\endcsname\relax
  \def\bibnamefont#1{#1}\fi
\expandafter\ifx\csname bibfnamefont\endcsname\relax
  \def\bibfnamefont#1{#1}\fi
\expandafter\ifx\csname citenamefont\endcsname\relax
  \def\citenamefont#1{#1}\fi
\expandafter\ifx\csname url\endcsname\relax
  \def\url#1{\texttt{#1}}\fi
\expandafter\ifx\csname urlprefix\endcsname\relax\def\urlprefix{URL }\fi
\providecommand{\bibinfo}[2]{#2}
\providecommand{\eprint}[2][]{\url{#2}}

\bibitem[{\citenamefont{Sachdev}(2011)}]{SS}
\bibinfo{author}{\bibfnamefont{S.}~\bibnamefont{Sachdev}},
  \emph{\bibinfo{title}{Quantum Phase Transitions}}
  (\bibinfo{publisher}{Cambridge University Press}, \bibinfo{year}{2011}).

\bibitem[{\citenamefont{Polkovnikov et~al.}(2011)\citenamefont{Polkovnikov,
  Sengupta, Silva, and Vengalattore}}]{PSSV11}
\bibinfo{author}{\bibfnamefont{A.}~\bibnamefont{Polkovnikov}},
  \bibinfo{author}{\bibfnamefont{K.}~\bibnamefont{Sengupta}},
  \bibinfo{author}{\bibfnamefont{A.}~\bibnamefont{Silva}}, \bibnamefont{and}
  \bibinfo{author}{\bibfnamefont{M.}~\bibnamefont{Vengalattore}},
  \bibinfo{journal}{Rev. Mod. Phys.} \textbf{\bibinfo{volume}{83}},
  \bibinfo{pages}{863} (\bibinfo{year}{2011}).

\bibitem[{\citenamefont{Calabrese et~al.}(2016)\citenamefont{Calabrese, Essler,
  and Mussardo}}]{CEM16}
\bibinfo{author}{\bibfnamefont{P.}~\bibnamefont{Calabrese}},
  \bibinfo{author}{\bibfnamefont{F.~H.~L.} \bibnamefont{Essler}},
  \bibnamefont{and} \bibinfo{author}{\bibfnamefont{G.}~\bibnamefont{Mussardo}},
  \bibinfo{journal}{J. Stat. Mech.} \bibinfo{eid}{064001}
  (\bibinfo{year}{2016}).

\bibitem[{\citenamefont{Heyl et~al.}(2013)\citenamefont{Heyl, Polkovnikov, and
  Kehrein}}]{HPK13}
\bibinfo{author}{\bibfnamefont{M.}~\bibnamefont{Heyl}},
  \bibinfo{author}{\bibfnamefont{A.}~\bibnamefont{Polkovnikov}},
  \bibnamefont{and} \bibinfo{author}{\bibfnamefont{S.}~\bibnamefont{Kehrein}},
  \bibinfo{journal}{Phys. Rev. Lett.} \textbf{\bibinfo{volume}{110}},
  \bibinfo{pages}{135704} (\bibinfo{year}{2013}).

\bibitem[{\citenamefont{Heyl}(2018)}]{Heyl17}
\bibinfo{author}{\bibfnamefont{M.}~\bibnamefont{Heyl}}, \bibinfo{journal}{Rep.
  Prog. Phys.} \textbf{\bibinfo{volume}{81}}, \bibinfo{pages}{054001}
  (\bibinfo{year}{2018}).

\bibitem[{\citenamefont{Jurcevic et~al.}(2017)\citenamefont{Jurcevic, Shen,
  Hauke, Maier, Brydges, Hempel, Lanyon, Heyl, Blatt, and Roos}}]{Jurcevic17}
\bibinfo{author}{\bibfnamefont{P.}~\bibnamefont{Jurcevic}},
  \bibinfo{author}{\bibfnamefont{H.}~\bibnamefont{Shen}},
  \bibinfo{author}{\bibfnamefont{P.}~\bibnamefont{Hauke}},
  \bibinfo{author}{\bibfnamefont{C.}~\bibnamefont{Maier}},
  \bibinfo{author}{\bibfnamefont{T.}~\bibnamefont{Brydges}},
  \bibinfo{author}{\bibfnamefont{C.}~\bibnamefont{Hempel}},
  \bibinfo{author}{\bibfnamefont{B.~P.} \bibnamefont{Lanyon}},
  \bibinfo{author}{\bibfnamefont{M.}~\bibnamefont{Heyl}},
  \bibinfo{author}{\bibfnamefont{R.}~\bibnamefont{Blatt}}, \bibnamefont{and}
  \bibinfo{author}{\bibfnamefont{C.~F.} \bibnamefont{Roos}},
  \bibinfo{journal}{Phys. Rev. Lett.} \textbf{\bibinfo{volume}{119}},
  \bibinfo{pages}{080501} (\bibinfo{year}{2017}).

\bibitem[{\citenamefont{Eckstein et~al.}(2009)\citenamefont{Eckstein, Kollar,
  and Werner}}]{EKW09}
\bibinfo{author}{\bibfnamefont{M.}~\bibnamefont{Eckstein}},
  \bibinfo{author}{\bibfnamefont{M.}~\bibnamefont{Kollar}}, \bibnamefont{and}
  \bibinfo{author}{\bibfnamefont{P.}~\bibnamefont{Werner}},
  \bibinfo{journal}{Phys. Rev. Lett.} \textbf{\bibinfo{volume}{103}},
  \bibinfo{pages}{056403} (\bibinfo{year}{2009}).

\bibitem[{\citenamefont{Schir\'o and Fabrizio}(2010)}]{SF10}
\bibinfo{author}{\bibfnamefont{M.}~\bibnamefont{Schir\'o}} \bibnamefont{and}
  \bibinfo{author}{\bibfnamefont{M.}~\bibnamefont{Fabrizio}},
  \bibinfo{journal}{Phys. Rev. Lett.} \textbf{\bibinfo{volume}{105}},
  \bibinfo{pages}{076401} (\bibinfo{year}{2010}).

\bibitem[{\citenamefont{Sciolla and Biroli}(2010)}]{SB10}
\bibinfo{author}{\bibfnamefont{B.}~\bibnamefont{Sciolla}} \bibnamefont{and}
  \bibinfo{author}{\bibfnamefont{G.}~\bibnamefont{Biroli}},
  \bibinfo{journal}{Phys. Rev. Lett.} \textbf{\bibinfo{volume}{105}},
  \bibinfo{pages}{220401} (\bibinfo{year}{2010}).

\bibitem[{\citenamefont{Garrahan and Lesanovsky}(2010)}]{GL10}
\bibinfo{author}{\bibfnamefont{J.~P.} \bibnamefont{Garrahan}} \bibnamefont{and}
  \bibinfo{author}{\bibfnamefont{I.}~\bibnamefont{Lesanovsky}},
  \bibinfo{journal}{Phys. Rev. Lett.} \textbf{\bibinfo{volume}{104}},
  \bibinfo{pages}{160601} (\bibinfo{year}{2010}).

\bibitem[{\citenamefont{Diehl et~al.}(2010)\citenamefont{Diehl, Tomadin,
  Micheli, Fazio, and Zoller}}]{DTMFZ10}
\bibinfo{author}{\bibfnamefont{S.}~\bibnamefont{Diehl}},
  \bibinfo{author}{\bibfnamefont{A.}~\bibnamefont{Tomadin}},
  \bibinfo{author}{\bibfnamefont{A.}~\bibnamefont{Micheli}},
  \bibinfo{author}{\bibfnamefont{R.}~\bibnamefont{Fazio}}, \bibnamefont{and}
  \bibinfo{author}{\bibfnamefont{P.}~\bibnamefont{Zoller}},
  \bibinfo{journal}{Phys. Rev. Lett.} \textbf{\bibinfo{volume}{105}},
  \bibinfo{pages}{015702} (\bibinfo{year}{2010}).

\bibitem[{\citenamefont{Halimeh and Zauner-Stauber}(2017)}]{HZ17}
\bibinfo{author}{\bibfnamefont{J.~C.} \bibnamefont{Halimeh}} \bibnamefont{and}
  \bibinfo{author}{\bibfnamefont{V.}~\bibnamefont{Zauner-Stauber}},
  \bibinfo{journal}{Phys. Rev. B} \textbf{\bibinfo{volume}{96}},
  \bibinfo{pages}{134427} (\bibinfo{year}{2017}).

\bibitem[{\citenamefont{\v{Z}unkovi\v{c}
  et~al.}(2018)\citenamefont{\v{Z}unkovi\v{c}, Heyl, Knap, and Silva}}]{ZHKS18}
\bibinfo{author}{\bibfnamefont{B.}~\bibnamefont{\v{Z}unkovi\v{c}}},
  \bibinfo{author}{\bibfnamefont{M.}~\bibnamefont{Heyl}},
  \bibinfo{author}{\bibfnamefont{M.}~\bibnamefont{Knap}}, \bibnamefont{and}
  \bibinfo{author}{\bibfnamefont{A.}~\bibnamefont{Silva}},
  \bibinfo{journal}{Phys. Rev. Lett.} \textbf{\bibinfo{volume}{120}},
  \bibinfo{pages}{130601} (\bibinfo{year}{2018}).

\bibitem[{\citenamefont{Prosen and Pi\v{z}orn}(2008)}]{PP08}
\bibinfo{author}{\bibfnamefont{T.}~\bibnamefont{Prosen}} \bibnamefont{and}
  \bibinfo{author}{\bibfnamefont{I.}~\bibnamefont{Pi\v{z}orn}},
  \bibinfo{journal}{Phys. Rev. Lett.} \textbf{\bibinfo{volume}{101}},
  \bibinfo{pages}{105701} (\bibinfo{year}{2008}).

\bibitem[{\citenamefont{Prosen and \v{Z}unkovi\v{c}}(2010)}]{PZ10}
\bibinfo{author}{\bibfnamefont{T.}~\bibnamefont{Prosen}} \bibnamefont{and}
  \bibinfo{author}{\bibfnamefont{B.}~\bibnamefont{\v{Z}unkovi\v{c}}},
  \bibinfo{journal}{New. J. Phys.} \textbf{\bibinfo{volume}{12}},
  \bibinfo{pages}{025016} (\bibinfo{year}{2010}).

\bibitem[{sm()}]{sm}
\bibinfo{note}{See Supplemental Material for details.}

\bibitem[{\citenamefont{Iorgov}(2011)}]{Iorgov11}
\bibinfo{author}{\bibfnamefont{N.}~\bibnamefont{Iorgov}}, \bibinfo{journal}{J.
  Phys. A: Math. Theor.} \textbf{\bibinfo{volume}{44}}, \bibinfo{pages}{335005}
  (\bibinfo{year}{2011}).

\bibitem[{\citenamefont{Iorgov and Lisovyy}(2011)}]{IL11}
\bibinfo{author}{\bibfnamefont{N.}~\bibnamefont{Iorgov}} \bibnamefont{and}
  \bibinfo{author}{\bibfnamefont{O.}~\bibnamefont{Lisovyy}},
  \bibinfo{journal}{J. Stat. Mech.} \bibinfo{eid}{P04011}
  (\bibinfo{year}{2011}).

\bibitem[{\citenamefont{Eisler et~al.}(2016)\citenamefont{Eisler, Maislinger,
  and Evertz}}]{EME16}
\bibinfo{author}{\bibfnamefont{V.}~\bibnamefont{Eisler}},
  \bibinfo{author}{\bibfnamefont{F.}~\bibnamefont{Maislinger}},
  \bibnamefont{and} \bibinfo{author}{\bibfnamefont{H.~G.}
  \bibnamefont{Evertz}}, \bibinfo{journal}{SciPost Phys.}
  \textbf{\bibinfo{volume}{1}}, \bibinfo{pages}{014} (\bibinfo{year}{2016}).

\bibitem[{\citenamefont{Sachdev and Young}(1997)}]{SY97}
\bibinfo{author}{\bibfnamefont{S.}~\bibnamefont{Sachdev}} \bibnamefont{and}
  \bibinfo{author}{\bibfnamefont{A.~P.} \bibnamefont{Young}},
  \bibinfo{journal}{Phys. Rev. Lett.} \textbf{\bibinfo{volume}{78}},
  \bibinfo{pages}{2220} (\bibinfo{year}{1997}).

\bibitem[{\citenamefont{Antal et~al.}(2008)\citenamefont{Antal, Krapivsky, and
  R{\'a}kos}}]{AKR08}
\bibinfo{author}{\bibfnamefont{T.}~\bibnamefont{Antal}},
  \bibinfo{author}{\bibfnamefont{P.~L.} \bibnamefont{Krapivsky}},
  \bibnamefont{and}
  \bibinfo{author}{\bibfnamefont{A.}~\bibnamefont{R{\'a}kos}},
  \bibinfo{journal}{Phys. Rev. E} \textbf{\bibinfo{volume}{78}},
  \bibinfo{pages}{061115} (\bibinfo{year}{2008}).

\bibitem[{\citenamefont{Rieger and Igl\'oi}(2011)}]{RI11}
\bibinfo{author}{\bibfnamefont{H.}~\bibnamefont{Rieger}} \bibnamefont{and}
  \bibinfo{author}{\bibfnamefont{F.}~\bibnamefont{Igl\'oi}},
  \bibinfo{journal}{Phys. Rev. B} \textbf{\bibinfo{volume}{84}},
  \bibinfo{pages}{165117} (\bibinfo{year}{2011}).

\bibitem[{\citenamefont{Kormos et~al.}(2018)\citenamefont{Kormos, Moca, and
  Zar\'and}}]{KMZ17}
\bibinfo{author}{\bibfnamefont{M.}~\bibnamefont{Kormos}},
  \bibinfo{author}{\bibfnamefont{C.~P.} \bibnamefont{Moca}}, \bibnamefont{and}
  \bibinfo{author}{\bibfnamefont{G.}~\bibnamefont{Zar\'and}},
  \bibinfo{journal}{Phys. Rev. E} \textbf{\bibinfo{volume}{98}},
  \bibinfo{pages}{032105} (\bibinfo{year}{2018}).

\bibitem[{\citenamefont{Eisler and R\'acz}(2013)}]{ER13}
\bibinfo{author}{\bibfnamefont{V.}~\bibnamefont{Eisler}} \bibnamefont{and}
  \bibinfo{author}{\bibfnamefont{Z.}~\bibnamefont{R\'acz}},
  \bibinfo{journal}{Phys. Rev. Lett.} \textbf{\bibinfo{volume}{110}},
  \bibinfo{pages}{060602} (\bibinfo{year}{2013}).

\bibitem[{\citenamefont{Viti et~al.}(2016)\citenamefont{Viti, \relax{J-M.}
  St\'ephan, Dubail, and Haque}}]{VSDH15}
\bibinfo{author}{\bibfnamefont{J.}~\bibnamefont{Viti}},
  \bibinfo{author}{\bibnamefont{\relax{J-M.} St\'ephan}},
  \bibinfo{author}{\bibfnamefont{J.}~\bibnamefont{Dubail}}, \bibnamefont{and}
  \bibinfo{author}{\bibfnamefont{M.}~\bibnamefont{Haque}},
  \bibinfo{journal}{Europhys. Lett.} \textbf{\bibinfo{volume}{115}},
  \bibinfo{pages}{40011} (\bibinfo{year}{2016}).

\bibitem[{\citenamefont{Allegra et~al.}(2016)\citenamefont{Allegra, Dubail,
  \relax{J-M.} St\'ephan, and Viti}}]{ADSV16}
\bibinfo{author}{\bibfnamefont{N.}~\bibnamefont{Allegra}},
  \bibinfo{author}{\bibfnamefont{J.}~\bibnamefont{Dubail}},
  \bibinfo{author}{\bibnamefont{\relax{J-M.} St\'ephan}}, \bibnamefont{and}
  \bibinfo{author}{\bibfnamefont{J.}~\bibnamefont{Viti}}, \bibinfo{journal}{J.
  Stat. Mech.} \bibinfo{eid}{053108} (\bibinfo{year}{2016}).

\bibitem[{\citenamefont{Perfetto and Gambassi}(2017)}]{PG17}
\bibinfo{author}{\bibfnamefont{G.}~\bibnamefont{Perfetto}} \bibnamefont{and}
  \bibinfo{author}{\bibfnamefont{A.}~\bibnamefont{Gambassi}},
  \bibinfo{journal}{Phys. Rev. E} \textbf{\bibinfo{volume}{96}},
  \bibinfo{pages}{012138} (\bibinfo{year}{2017}).

\bibitem[{\citenamefont{Kormos}(2017)}]{Kormos17}
\bibinfo{author}{\bibfnamefont{M.}~\bibnamefont{Kormos}},
  \bibinfo{journal}{SciPost Phys.} \textbf{\bibinfo{volume}{3}},
  \bibinfo{pages}{020} (\bibinfo{year}{2017}).

\bibitem[{\citenamefont{Tracy and Widom}(1994)}]{TW94}
\bibinfo{author}{\bibfnamefont{C.~A.} \bibnamefont{Tracy}} \bibnamefont{and}
  \bibinfo{author}{\bibfnamefont{H.}~\bibnamefont{Widom}},
  \bibinfo{journal}{Commun. Math. Phys.} \textbf{\bibinfo{volume}{159}},
  \bibinfo{pages}{151} (\bibinfo{year}{1994}).

\bibitem[{\citenamefont{Fagotti}(2017)}]{Fagotti17}
\bibinfo{author}{\bibfnamefont{M.}~\bibnamefont{Fagotti}},
  \bibinfo{journal}{Phys. Rev. B} \textbf{\bibinfo{volume}{96}},
  \bibinfo{pages}{220302(R)} (\bibinfo{year}{2017}).

\bibitem[{\citenamefont{Vidal et~al.}(2003)\citenamefont{Vidal, Latorre, Rico,
  and Kitaev}}]{Vidal03}
\bibinfo{author}{\bibfnamefont{G.}~\bibnamefont{Vidal}},
  \bibinfo{author}{\bibfnamefont{J.~I.} \bibnamefont{Latorre}},
  \bibinfo{author}{\bibfnamefont{E.}~\bibnamefont{Rico}}, \bibnamefont{and}
  \bibinfo{author}{\bibfnamefont{A.}~\bibnamefont{Kitaev}},
  \bibinfo{journal}{Phys. Rev. Lett.} \textbf{\bibinfo{volume}{90}},
  \bibinfo{pages}{227902} (\bibinfo{year}{2003}).

\bibitem[{\citenamefont{Peschel and Eisler}(2009)}]{PE09}
\bibinfo{author}{\bibfnamefont{I.}~\bibnamefont{Peschel}} \bibnamefont{and}
  \bibinfo{author}{\bibfnamefont{V.}~\bibnamefont{Eisler}},
  \bibinfo{journal}{J. Phys. A: Math. Theor.} \textbf{\bibinfo{volume}{42}},
  \bibinfo{pages}{504003} (\bibinfo{year}{2009}).

\bibitem[{\citenamefont{Fagotti and Essler}(2013)}]{FE13}
\bibinfo{author}{\bibfnamefont{M.}~\bibnamefont{Fagotti}} \bibnamefont{and}
  \bibinfo{author}{\bibfnamefont{F.~H.~L.} \bibnamefont{Essler}},
  \bibinfo{journal}{Phys. Rev. B} \textbf{\bibinfo{volume}{87}},
  \bibinfo{pages}{245107} (\bibinfo{year}{2013}).

\bibitem[{\citenamefont{Caputa and Rams}(2017)}]{CR17}
\bibinfo{author}{\bibfnamefont{P.}~\bibnamefont{Caputa}} \bibnamefont{and}
  \bibinfo{author}{\bibfnamefont{M.~M.} \bibnamefont{Rams}},
  \bibinfo{journal}{J. Phys. A: Math. Theor.} \textbf{\bibinfo{volume}{50}},
  \bibinfo{pages}{055002} (\bibinfo{year}{2017}).

\bibitem[{\citenamefont{Lanford and Robinson}(1968)}]{LR68}
\bibinfo{author}{\bibfnamefont{O.~E.} \bibnamefont{Lanford}} \bibnamefont{and}
  \bibinfo{author}{\bibfnamefont{D.~W.} \bibnamefont{Robinson}},
  \bibinfo{journal}{J. Math. Phys} \textbf{\bibinfo{volume}{9}},
  \bibinfo{pages}{1120} (\bibinfo{year}{1968}).

\bibitem[{\citenamefont{Wehrl}(1978)}]{Wehrl78}
\bibinfo{author}{\bibfnamefont{A.}~\bibnamefont{Wehrl}}, \bibinfo{journal}{Rev.
  Mod. Phys.} \textbf{\bibinfo{volume}{50}}, \bibinfo{pages}{221}
  (\bibinfo{year}{1978}).

\bibitem[{\citenamefont{Schollw{\"o}ck}(2011)}]{Schollwoeck11}
\bibinfo{author}{\bibfnamefont{U.}~\bibnamefont{Schollw{\"o}ck}},
  \bibinfo{journal}{Annals of Physics} \textbf{\bibinfo{volume}{326}},
  \bibinfo{pages}{96} (\bibinfo{year}{2011}).

\bibitem[{\citenamefont{Franchini}(2017)}]{FF}
\bibinfo{author}{\bibfnamefont{F.}~\bibnamefont{Franchini}},
  \emph{\bibinfo{title}{An Introduction to Integrable Techniques for
  One-Dimensional Quantum Systems}}, vol. \bibinfo{volume}{940} of
  \emph{\bibinfo{series}{Lecture Notes in Physics}}
  (\bibinfo{publisher}{Springer}, \bibinfo{year}{2017}).

\bibitem[{\citenamefont{Eisler and Maislinger}()}]{EMup}
\bibinfo{author}{\bibfnamefont{V.}~\bibnamefont{Eisler}} \bibnamefont{and}
  \bibinfo{author}{\bibfnamefont{F.}~\bibnamefont{Maislinger}},
  \bibinfo{note}{to be published}.

\bibitem[{\citenamefont{Bertini et~al.}(2016)\citenamefont{Bertini, Collura,
  \relax{De Nardis}, and Fagotti}}]{BCDNF16}
\bibinfo{author}{\bibfnamefont{B.}~\bibnamefont{Bertini}},
  \bibinfo{author}{\bibfnamefont{M.}~\bibnamefont{Collura}},
  \bibinfo{author}{\bibfnamefont{J.}~\bibnamefont{\relax{De Nardis}}},
  \bibnamefont{and} \bibinfo{author}{\bibfnamefont{M.}~\bibnamefont{Fagotti}},
  \bibinfo{journal}{Phys. Rev. Lett.} \textbf{\bibinfo{volume}{117}},
  \bibinfo{pages}{207201} (\bibinfo{year}{2016}).

\bibitem[{\citenamefont{Castro-Alvaredo
  et~al.}(2016)\citenamefont{Castro-Alvaredo, Doyon, and Yoshimura}}]{CADY16}
\bibinfo{author}{\bibfnamefont{O.~A.} \bibnamefont{Castro-Alvaredo}},
  \bibinfo{author}{\bibfnamefont{B.}~\bibnamefont{Doyon}}, \bibnamefont{and}
  \bibinfo{author}{\bibfnamefont{T.}~\bibnamefont{Yoshimura}},
  \bibinfo{journal}{Phys. Rev. X} \textbf{\bibinfo{volume}{6}},
  \bibinfo{pages}{041065} (\bibinfo{year}{2016}).

\bibitem[{\citenamefont{Piroli et~al.}(2017)\citenamefont{Piroli, \relax{De
  Nardis}, Collura, Bertini, and Fagotti}}]{PDNCBF17}
\bibinfo{author}{\bibfnamefont{L.}~\bibnamefont{Piroli}},
  \bibinfo{author}{\bibfnamefont{J.}~\bibnamefont{\relax{De Nardis}}},
  \bibinfo{author}{\bibfnamefont{M.}~\bibnamefont{Collura}},
  \bibinfo{author}{\bibfnamefont{B.}~\bibnamefont{Bertini}}, \bibnamefont{and}
  \bibinfo{author}{\bibfnamefont{M.}~\bibnamefont{Fagotti}},
  \bibinfo{journal}{Phys. Rev. B} \textbf{\bibinfo{volume}{96}},
  \bibinfo{pages}{115124} (\bibinfo{year}{2017}).

\bibitem[{\citenamefont{Collura et~al.}(2018)\citenamefont{Collura, \relax{De
  Luca}, and Viti}}]{CDLV18}
\bibinfo{author}{\bibfnamefont{M.}~\bibnamefont{Collura}},
  \bibinfo{author}{\bibfnamefont{A.}~\bibnamefont{\relax{De Luca}}},
  \bibnamefont{and} \bibinfo{author}{\bibfnamefont{J.}~\bibnamefont{Viti}},
  \bibinfo{journal}{Phys. Rev. B} \textbf{\bibinfo{volume}{97}},
  \bibinfo{pages}{081111(R)} (\bibinfo{year}{2018}).

\bibitem[{\citenamefont{Nozaki et~al.}(2014)\citenamefont{Nozaki, Numasawa, and
  Takayanagi}}]{NNT14}
\bibinfo{author}{\bibfnamefont{M.}~\bibnamefont{Nozaki}},
  \bibinfo{author}{\bibfnamefont{T.}~\bibnamefont{Numasawa}}, \bibnamefont{and}
  \bibinfo{author}{\bibfnamefont{T.}~\bibnamefont{Takayanagi}},
  \bibinfo{journal}{Phys. Rev. Lett.} \textbf{\bibinfo{volume}{112}},
  \bibinfo{pages}{111602} (\bibinfo{year}{2014}).

\end{thebibliography}

\clearpage
\onecolumngrid
\begin{center}
\textbf{\large Supplemental Material: \\ Hydrodynamical phase transition for domain-wall melting in the XY chain}
\end{center}
\setcounter{equation}{0}
\setcounter{figure}{0}
\setcounter{table}{0}
\setcounter{page}{1}
\makeatletter
\renewcommand{\theequation}{S\arabic{equation}}
\renewcommand{\thefigure}{S\arabic{figure}}

\section{Fermionization of XY Hamiltonian}

In order to obtain the many-body eigenstates of the XY chain,
it is useful to consider periodic $H_+$ or antiperiodic $H_-$ chains,
instead of the open one in Eq. \eqref{hxy}. These are given by
\eq{
H_s=- \frac{1}{2}\sum_{n=1}^{N} \left( \frac{1+\gamma}{2}\sigma_n^x 
\sigma_{n+1}^x+\frac{1-\gamma}{2}\sigma_n^y \sigma_{n+1}^y
+h \sigma_n^z\right),
}
where the boundary conditions are $\sigma^x_{N+1}=s \sigma^x_1$
and $\sigma^y_{N+1}=s \sigma^y_1$ for $s=\pm$.
Since $H_s$ commutes with the parity $P$, it can be written in a block-diagonal form
\eq{
H_s = \frac{1-sP}{2}H_{\mathrm{R}} + \frac{1+sP}{2}H_{\mathrm{NS}}, \qquad
P = \prod_{n=1}^{N} \sigma^z_n \, .
}
The parity subspaces are the Ramond (R) and Neveu-Schwarz (NS) sectors, defining
two different Hamiltonians.
In terms of Majorana operators, obtained via the Jordan-Wigner transformation \eqref{maj},
both of them can be brought into the quadratic form 
\eq{
H_{\mathrm{R/NS}}=\frac{i}{2}\sum_{j=1}^{N} \left(
\frac{1+\gamma}{2} a_{2j}a_{2j+1}
-\frac{1-\gamma}{2} a_{2j-1}a_{2j+2}
+h a_{2j-1}a_{2j} \right),\qquad
}
where the two Hamiltonians differ only in the boundary conditions
$a_{2N+1}=\pm a_1$ and $a_{2N+2}=\pm a_2$ being periodic
for R and antiperiodic for the NS sector. Each sector can be simultaneously
diagonalized by a joint Fourier and Bogoliubov transformation
\eq{
a_{2j-1}=\frac{1}{\sqrt{N}} \sum_{q \in \mathrm{R/NS}}
\ee^{-iq(j-1)} \ee^{i\theta_q/2} (b_q^\dag + b_{-q}), \qquad
a_{2j}=\frac{-i}{\sqrt{N}} \sum_{q \in \mathrm{R/NS}} \ee^{-iqj}
\ee^{-i\theta_q/2} (b_q^\dag - b_{-q}),
\label{ab}}
where the allowed values of the momenta are $q_k=\frac{2\pi}{N}k$ for R and
$q_k=\frac{2\pi}{N}(k+1/2)$ for NS, respectively, with $k=-N/2,\dots,N/2-1$.
Note that the site index $j$ in the Fourier transformation is shifted by one for odd
Majorana operators. This is a dual representation in terms of which
the Bogoliubov angle must satisfy
\eq{
\ee^{i(\theta_q+q)} = \frac{\cos q -h + i \gamma \sin q}{\epsilon_q}, \qquad
\epsilon_q = \sqrt{(\cos q-h)^2+\gamma^2 \sin^2 q} \, .
}
In fact, the above definition ensures that $\theta_q$ is a continuous and smooth
function in its full domain $q \in \left[-\pi,\pi\right]$, for arbitrary parameters
$0<\gamma\le 1$ and $0 \le h<1$ in the ferromagnetic phase.
The diagonal form of the Hamiltonian and its many-particle eigenstates then read
\eq{
H_{\mathrm{R/NS}} = \sum_{q \in \mathrm{R/NS}}
{\epsilon_q} b^\dag_q b_q + \mathrm{const}, \qquad
|q_1,q_2,\dots,q_m\rangle_{\mathrm{R/NS}} = 
\prod_{i=1}^{m} b^\dag_{q_i} |0\rangle_{\mathrm{R/NS}} \, .
}
Finally, it should be pointed out that the boundary condition on the spins
selects the parity of the many-particle basis: $m=2\ell$ is even for the spin-periodic
Hamiltonian $H_+$, and $m=2\ell+1$ is odd for the spin-antiperiodic one $H_-$.

\section{Form factor approach}

To calculate the time evolution of the magnetization, one also needs
the corresponding form factors of the $\sigma^x$ operator.
In fact, it is more convenient to consider the matrix elements normalized
by the equilibrium magnetization, which in the large $N$ limit reads
\cite{Iorgov11,IL11}
\eq{
\rl{p}\hat{\mathcal{M}}_n\nsr{q} =
\frac{\rl{p} \sigma^x_n \nsr{q}}{\rl{0} \sigma^x_n \nsr{0}} = -\frac{i}{N}
\frac{\cosh\frac{\Delta_p-\Delta_q}{2}\sinh\frac{\Delta_p+\Delta_q}{2}}
{\sqrt{\sinh \Delta_p \sinh\Delta_q}}
\frac{\ee^{i(n-1/2)(q-p)}}{\sin \frac{q-p}{2}}.
\label{ff2}}
The above definition of the form factors is well-suited for the
parameter regime $\sqrt{1-\gamma^2}<h<1$, i.e. in the non-oscillatory
ferromagnetic phase \cite{FF}, where the auxiliary parameter $\Delta_q$ is defined via
\eq{
\sinh \Delta_q = 
\frac{\sqrt{1-\gamma^2}}{\gamma \sqrt{\gamma^2+h^2-1}} \epsilon_q .
\label{sinh1}}
In the oscillatory phase $0<h< \sqrt{1-\gamma^2}$ the form factors
can be obtained by analytic continuation \cite{Iorgov11},
i.e. by introducing the variable $\tilde \Delta_q = \Delta_q + i\pi/2$.
In fact, the form-factor formula \eqref{ff2} can even be further simplified
by making use of the identity
\eq{
\cosh\frac{\Delta_p-\Delta_q}{2}\sinh\frac{\Delta_p+\Delta_q}{2}=
\frac{1}{2}(\sinh \Delta_p+\sinh \Delta_q).
\label{sinh2}}
Substituting \eqref{sinh1} and \eqref{sinh2} into \eqref{ff2},
one obtains immediately Eq. \eqref{ff}. Using these form factors and
taking the thermodynamic limit, Eq. \eqref{mnt} for the magnetization
can be written out as a double integral
\eq{
\mnt=
\ip \int_{-\pi} ^{\pi} \frac{\dd p}{2\pi} \int_{-\pi}^{\pi} \frac{\dd q}{2\pi}
\frac{\epsilon_p + \epsilon_q}{2\sqrt{\epsilon_p \epsilon_q}}
 \frac{\ee^{i(n-n_0+1/2)(q-p)}}{\sin\frac{q-p}{2}}
\ee^{i(\theta_q - \theta_p)/2} \ee^{i(\epsilon_p-\epsilon_q)t} \, .
\label{mnint}}
Using the properties $\epsilon_{-q}=\epsilon_q$ and
$\theta_{-q}=-\theta_{q}$, the above expression can be written as
$\mnt = \mathcal{M}^e_n(t)+\mathcal{M}^o_n(t)$ with only two
nonvanishing contributions
\eq{
\begin{split}
\mathcal{M}^e_n(t)=
\int_{-\pi} ^{\pi} \frac{\dd p}{2\pi} \int_{-\pi}^{\pi} \frac{\dd q}{2\pi}
\frac{\epsilon_p + \epsilon_q}{2\sqrt{\epsilon_p \epsilon_q}}
 \frac{\cos\left[(n-n_0+1/2)(q-p)\right]}{\sin\frac{q-p}{2}}
\sin\frac{\theta_q - \theta_p}{2} \cos(\epsilon_p-\epsilon_q)t \, ,\\
\mathcal{M}^o_n(t)=
\int_{-\pi} ^{\pi} \frac{\dd p}{2\pi} \int_{-\pi}^{\pi} \frac{\dd q}{2\pi}
\frac{\epsilon_p + \epsilon_q}{2\sqrt{\epsilon_p \epsilon_q}}
 \frac{\sin\left[(n-n_0+1/2)(q-p)\right]}{\sin\frac{q-p}{2}}
\cos\frac{\theta_q - \theta_p}{2} \cos(\epsilon_p-\epsilon_q)t \, .
\end{split}
\label{mneoint}}
Hence the magnetization is the sum of an even and an odd function
$\mathcal{M}^{e,o}_n(t) = \pm \mathcal{M}^{e,o}_{2n_0-1-n}(t)$ under reflections
with respect to the initial domain wall position.
Note that, in general, the even term has a contribution of much smaller
magnitude, and it vanishes completely in the hydrodynamic scaling limit.
Moreover, in the limit $\gamma=1$ of a transverse Ising chain, the even
part $\mathcal{M}^{e}_n(t) = 0$ vanishes identically even for finite times.

The normalized correlation functions
$\mathcal{C}_{m,n}(t)= \nsl{\phi_t} \hat{\mathcal{M}}_m \hat{\mathcal{M}}_n \nsr{\phi_t}$
can also be studied through the form factor approach. The standard trick is to
insert an identity between the two operators, written in terms of the eigenbasis
\eq{
\identity = \sum_{p} |p\rangle \langle p | +
\sum_{p_1,p_2} |p_1,p_2\rangle \langle p_1,p_2 | +
\sum_{p_1,p_2,p_3} |p_1,p_2,p_3\rangle \langle p_1,p_2,p_3 | + \dots
}
Thus, in contrast to the magnetization which could be exactly evaluated
using only single-particle form factors, the situation for the correlations is much more
complicated as an infinite series of many-particle matrix elements appear.
Nevertheless, it is reasonable to expect that the dominant contribution to the
correlations still comes from the single-particle sector. Hence, we will consider this
approximate expression, given by \eqref{cmnt} in the main text, which for $N\to\infty$
can be converted into the integral form
\eq{
\mathcal{C}_{m,n}(t) \simeq
\int \frac{\dd q_1}{2\pi}\int \frac{\dd q_2}{2\pi}
\ee^{-i(\theta_{q_1}-\theta_{q_2})/2}\ee^{i(\epsilon_{q_1} -\epsilon_{q_2})t}
\int \frac{\dd p}{2\pi} \frac{\epsilon_p + \epsilon_{q_1}}{2\sqrt{\epsilon_p \epsilon_{q_1}}}
\frac{\epsilon_p + \epsilon_{q_2}}{2\sqrt{\epsilon_p \epsilon_{q_2}}}
\frac{\ee^{-i(m-n_0+1/2)(q_1-p)}}{\sin \frac{q_1-p}{2}}\frac{\ee^{i(n-n_0+1/2)(q_2-p)}}{\sin \frac{q_2-p}{2}} \, .
\label{cmnint}}

\section{Stationary phase calculations}

The profiles in the hydrodynamic scaling limit can be obtained by stationary phase arguments,
and their derivation closely follows the lines of Refs. \cite{VSDH15,ADSV16,PG17,Kormos17}.
Let us consider first the magnetization as given by Eq. \eqref{mnint}. In the limit $n-n_0 \gg 1$ and
$t\gg 1$, the integrand is highly oscillatory and thus the main contribution comes from around the
points $q_s$ where the stationarity condition is satisfied
\eq{
v_{q_s}t = n-n_0+1/2+\theta'_{q_s}/2, \qquad
v_q = \frac{\dd \epsilon_q}{\dd q}.
}
The stationary phase condition for the integral over $p$ is exactly the same.
Moreover, the integrand has a pole at $p=q$ which suggests the change of variables
$Q=q-p$ and $P=(q+p)/2$. In the new variables, the stationarity condition is $Q_s=0$
for arbitrary values of $P$. One shall thus expand the integrand in \eqref{mnint}
around $Q=0$, setting
\eq{
\frac{\epsilon_p + \epsilon_q}{2\sqrt{\epsilon_p \epsilon_q}} \approx 1, \qquad
\sin \frac{q-p}{2} \approx \frac{Q}{2},
}
to arrive at
\eq{
2 \, \rp \int_{-\pi}^{\pi} \frac{\dd P}{2\pi}\int_{-\infty}^{\infty} \frac{\dd Q}{2\pi i}
\frac{\ee^{i(n-n_0+1/2+\theta'_{P}-v_{P}t)Q}}{Q}.
}
%
To carry out the integration around the pole, we use a formal identity in complex analysis
as well as the integral representation of the Heaviside theta function
\eq{
\frac{1}{Q} = i\pi \delta(Q) + \lim_{\delta\to0} \frac{1}{Q+i\delta}, \qquad
\Theta(x) = -\lim_{\delta \to 0} \int_{-\infty}^{\infty} \frac{\dd Q}{2\pi i}
\frac{\ee^{-iQx}}{Q+i\delta}.
\label{ht}
}
In the hydrodynamic regime one can neglect the term $\theta'_{P}$ and introduce
the scaling variable $\nu=(n-n_0+1/2)/t$, which brings us to the result
\eqref{mntsc} in the main text.

In general, the hydrodynamic profile is found by solving the equation $v_q=\nu$.
Special attention is needed around the maximum $v_{q_*}=v_{max}$ of the velocities,
where the solutions coalesce at momentum $q_*$. To get the fine structure of the
front edge, one has to expand the dispersion around $q_*$ as
\eq{
\epsilon_q \approx \epsilon_{q_*} + v_{q_*} (q-q_*) +
\frac{\epsilon'''_{q_*}}{6} (q- q_*)^3.
\label{epsqmax}}
Furthermore, one can introduce the following rescaled variables
\eq{
X = (n-n_0+1/2+\theta'_{q_*}/2-v_{q_*}t) \left(\frac{-2}{\epsilon'''_{q_*}t}\right)^{1/3}, \quad
Q = \left(\frac{-2}{\epsilon'''_{q_*}t}\right)^{-1/3} (q-q_*), \quad
P = \left(\frac{-2}{\epsilon'''_{q_*}t}\right)^{-1/3} (p-q_*).
\label{XQP}}
Substituting \eqref{epsqmax} and \eqref{XQP} into \eqref{mnt}, one arrives at the following integral
\eq{
\left(\frac{-2}{\epsilon'''_{q_*}t}\right)^{1/3}
\ip \int \frac{\dd P}{2\pi} \int \frac{\dd Q}{2\pi}
\frac{\ee^{iX(Q-P)} \ee^{i(Q^3-P^3)/3}}{(Q-P)/2}.
}
Using the integral representation of the Airy kernel
\eq{
K(X,Y) = \lim_{\delta\to0} \int \frac{\dd P}{2\pi} \int \frac{\dd Q}{2\pi}
\frac{\ee^{-iXP} \ee^{-iP^3/3} \ee^{iYQ} \ee^{iQ^3/3}}{i(P-Q-i\delta)}=
\frac{\Ai(X)\Ai'(Y)-\Ai'(X)\Ai(Y)}{X-Y},
\label{kxy}}
one recovers \eqref{medge} of the main text, where the diagonal terms
of the Airy kernel \cite{TW94} can be obtained as
\eq{
\rho(X)=\lim_{Y \to X} K(X,Y) = \left[ \Ai'(X) \right]^2 - X\Ai^2(X).
}

The stationary phase calculation for the approximation of the correlation function in \eqref{cmnint}
is very similar to that for the magnetization. Indeed, introducing the new set of variables
\eq{
Q_1 = q_1-p, \qquad
Q_2 = q_2-p, \qquad
P = \frac{q_1+p}{2},
}
and expanding around $Q_1=0$ and $Q_2=0$, one obtains
\eq{
\mathcal{C}_{m,n}(t) \simeq
4\int \frac{\dd P}{2\pi}
\int \frac{\dd Q_1}{2\pi}
\frac{\ee^{-i(m-n_0+1/2 + \theta'_P -v_Pt)Q_1}}{Q_1}
\int \frac{\dd Q_2}{2\pi}
\frac{\ee^{i(n-n_0+1/2 + \theta'_P -v_Pt)Q_2}}{Q_2}.
\label{cxint2}}
Applying \eqref{ht} in both the $Q_1$ and $Q_2$ integrals, the result can
again be written with the help of step functions. Using the property
$\Theta(x-x_1)\Theta(x-x_2)=\Theta(x-\max(x_1,x_2))$, and introducing the scaling
variable $\mu =(m-n_0+1/2)/t$ analogously to $\nu$, one arrives at
\eq{
\mathcal{C}_{m,n}(t) \simeq
1-2\int_{-\pi}^{\pi} \frac{\dd P}{2\pi}\Theta(v_{P} - \mu)+2\int_{-\pi}^{\pi} \frac{\dd P}{2\pi}\Theta(v_{P} - \nu),
}
where we assumed $\mu<\nu$. Finally, the difference of the step functions
can also be rewritten as a product as in \eqref{cmnt2}.

\end{document}